\documentstyle[12pt]{article}
\begin{document}

\title{
  Study of Superdeformation in Non-rotating States
using the Skyrme-Hartree-Fock Method
}

\author{
  Satoshi\ Takahara\thanks{Corresponding author. E-mail
  staka@hep1.c.u-tokyo.ac.jp}, Naoki\ Tajima and Naoki\ Onishi\\
  {\it Institute of Physics, Graduate School of Arts and Sciences,}\\
  {\it University of Tokyo, Komaba, 153-8902, Japan}
}

\date{}

\maketitle

\begin{abstract}
The superdeformation (SD) in non-rotating states is studied with the
HF+BCS method using the Skyrme interaction.
In applying the BCS theory, the seniority pairing force is employed,
of which strengths are determined in order to reproduce the empirical
pairing gap formula, $\bar{\Delta} = 12 A^{-1/2}$, through a smooth
level density obtained in the Thomas-Fermi approximation.
Properties of superdeformation are investigated by calculating
potential energy surfaces (PES) for various sets of the pairing force
strengths and the Skyrme force parameter for $^{194}$Hg and $^{236,
  238}$U.
The best results are obtained using both the SkM$^{\ast}$ force and
the pairing force strength determined in this paper.
By making use of this set of forces, a systematic calculation of SD
states is carried out extensively for even-even nuclei for $20 \le Z
\le 82$.
From our calculation, the barriers preventing the decay into the
normally deformed states are about twice as high as those predicted by
Krieger et al.\, who used the same Skyrme interaction but a pairing
force stronger than ours.  The differences of the present results from
the Nilsson-Strutinsky calculation are analyzed.

\vspace{.5cm}



\end{abstract}

\section{Introduction} \label{sect:intro}

In the last decade, superdeformed (SD) bands have been studied
extensively both experimentally and theoretically.
Since the first observation of SD bands was made in $^{152}$Dy nucleus
in 1986\cite{TW86},
they have been found in four mass regions, i.e.,
$A \sim 80$\cite{Ba95}, $130, 150$ and $190$\cite{NT88,Ab90,JK91}.
%
%
The theoretical studies of SD have helped to clarify
understanding of nuclear deformation.
Indeed the present authors carried out an extensive investigation
of normal deformations using Skyrme-Hartree-Fock method and reproduced
the deformation for almost all even-even nuclei\cite{TTO96}.
Therefore we thought it would be worthwhile to use the method
to calculate SD.

These SD bands have been found only at high spins so far,
mainly because of the experimental method of formation of SD bands.
It is expected that they may be present at zero spin,
like fission isomers in actinide nuclei.
It may require elaborate calculations to search for SD states in the
two-dimensional space of angular momentum and quadrupole moment.
Therefore, as a first step, we tried to study SD in non-rotating
states and did not crank the mean field.
%
%
%

In this paper, we looked for SD in non-rotating states using
numerical calculations of potential energy surfaces (PES) employing
the Hartree-Fock(HF) with the Skyrme interaction together with
the BCS method using the seniority force.
A characteristic feature of our calculations is that the single-particle
wave functions are expressed in a three-dimensional Cartesian-mesh
representation as described in Ref.\ \cite{BFH85}.
This representation enables one to obtain wave functions for various
shapes in a single framework, without any prejudice about shape,
i.e., one need not optimize the deformation of the basis
(like the three oscillator frequencies $\omega_{\kappa}$
of the anisotropic harmonic-oscillator basis) for each solution.
Solving the mean-field equations in this representation requires,
however, a large amount of computation which can be accomplished only
with supercomputers.

In this paper we  utilize an HF+BCS code {\sl EV8}\cite{BFH85},
in which the mean-field potential is assumed to be symmetric for
reflections in the $xy$-, $yz$-, and $zx$-planes
(the point group D$_{\rm 2h}$).
We set the mesh size to 1 fm, while the size of the box is
$15\times15\times16$ fm.  We put an octant of a nucleus at a corner of
the box. For the BCS part, we employ the seniority force.

In sect.~\ref{sect:hg}, we compare various parameter sets of the
Skyrme force to see how accurately they can reproduce the excitation
energy of the SD band head of $^{194}$Hg.  We explain our method to
determine the pairing force strengths.  We also show how strongly the
pairing correlation affects the properties of SD states.  In
sect.~\ref{sect:fi}, we study the fission isomers in $^{236,238}$U.
In sect.~\ref{sect:syshf}, we systematically search large-deformation
solutions employing the SkM$^{\ast}$ force. We compare the results
with those obtained by Krieger et al.\ \cite{KBW92}, who used the same
Skyrme force but stronger pairing force strengths.  In
sect.~\ref{sect:ns}, we also analyse the results obtained with the
Nilsson-Strutinsky method.  Section \ref{sect:sum} is a summary of
the present paper.

\section{Superdeformation in $^{194}$Hg} \label{sect:hg}

Recently, Khoo et al.\ \cite{Kh96} determined the excitation energies
and the spins of a SD band in $^{194}$Hg down to $I^{\pi}=8^{+}$.
By extrapolating the spectrum to $I=0$, they could predict reliably
the excitation energy of the band head to be 6.017 MeV.

In Table~\ref{tab:hg194}, theoretical predictions (listed in
Ref.\ \cite{Kh96}) are shown to be compared with the extrapolated
value. One can see that all of the predictions are either lower or
higher than the extrapolated value by at least 0.9 MeV.


Fig.\ \ref{fig:hg194} presents the results of our HF+BCS calculations
using nine sets of the Skyrme force parameters.
We draw PES curves as functions of the quadrupole deformation
parameter $\delta$ defined as
\begin{equation} \label{eq:delta}
  \delta \equiv \frac{3 \langle \hat{Q}_z\rangle}{4 \langle\hat{r}^2
    \rangle},
\end{equation}
where $\hat{Q}_z$ is the axially symmetric mass quadrupole moment,
$\hat{Q}_z$ $\equiv$ $2\hat{z}^2$ $-\hat{x}^2$ $-\hat{y}^2$, and
$\hat{r}^2$ is the squared mass radius, $\hat{r}^2$ $\equiv$
$\hat{x}^2$ $+\hat{y}^2$ $+\hat{z}^2$.  The excitation energy
$E^{\ast}$ of the SD minimum at $\delta \sim 0.5$ measured from the
ground-state minimum at $\delta \sim -0.1$ is given in parentheses in
units of MeV for each force.

The left-hand portion of Fig.\ \ref{fig:hg194} shows the results with
five widely-used Skyrme interactions; SIII\cite{BFG75},
SkSC4\cite{APD92}, SkM$^{\ast}$\cite{BQB82}, SkP\cite{DFT84}, and
SGII\cite{GS81}.  One can see that the SkM$^{\ast}$ as well as the SkP
forces are the best ones to reproduce the experimental value of
$E^{\ast}$. On the other hand, the SIII and the SkSC4 forces make the
nucleus too stiff against deformation, while the SGII too soft.  From
a macroscopic point of view, the softness toward deformation is
determined by the smallness of the surface energy coefficient $a_{\rm
  s}$\cite{BQB82} specific to each force. The SkM$^{\ast}$ is a force
adjusted so as to reproduce the fission barrier height of $^{240}$Pu
and thus is expected to have the correct surface energy coefficient.

The right-hand portion of Fig.\ \ref{fig:hg194} gives the results for
four new parameter sets, SkI2 - SkI5\cite{RF94}.  These forces have
different features involving the density dependence of the spin-orbit
potential.
The standard Skyrme interactions, such as SkI2 and SkI5, produce
spin-orbit potentials for neutrons which are proportional to the
gradient of $\rho + \rho_{\rm n}$, where $\rho$ ($\rho_{\rm n}$)
denotes the nucleon (neutron) density.  On the other hand, the
spin-orbit potentials produced by SkI3 and SkI4 are proportional to
the gradient of $\rho$ and roughly $ \rho - \rho_{\rm n}$(=$\rho_{\rm
  p}$), respectively.
The parameters of SkI2 - SkI4 were determined by fitting to the same
set of nuclear properties.  A slightly different set of the quantities
was adjusted to determine SkI5.  Among these four forces, SkI3
reproduces the experimental value of $E^{\ast}$ very accurately.  On
the other hand, SkI2 and SkI5 (i.e., standard Skyrme forces) make the
nucleus too soft against deformations while SkI4 too stiff.  This fact
is consistent with the relativistic mean-field theory, whose
``spin-orbit potential'' has a similar density dependence\cite{RF94}
to that of SkI3.


We now discuss the effects of pairing.  Using the same Skyrme
interaction SkM$^{\ast}$, Krieger's calculation\cite{KBW92} and ours
give different results, which originate in the difference of the
strengths of the seniority interaction used to describe the pairing
correlation.

Krieger et al.\ used the pairing force strengths $G_{\tau}$ ($\tau$=n
for neutrons and $\tau$=p for protons) given by the following formula,
\begin{equation} \label{eq:g}
  G_{\rm p}=\frac{17.5}{11+Z}, \;\;\;
  G_{\rm n}=\frac{16.5}{11+N},  \;\; [{\rm MeV}].
\end{equation}
On the other hand, we determined the strengths $G_{\tau}$
such that the so-called classical empirical formula of
the average pairing gap,
\begin{equation}
  \bar{\Delta}_{\tau} = \frac{12}{\sqrt{A}} \;\;\; [{\rm MeV}],
\end{equation}
is reproduced for shell-effect-averaged level density obtained in the
Thomas-Fermi approximation.  The pairing-active space is the same in
both the Krieger et al.\ treatment and ours: single-particle levels
below ``the Fermi level plus 5 MeV'' were taken into account in the
BCS calculation in both cases.  See Ref.\ \cite{TTO96} for details of
our calculations.

%
The effects of the change in the pairing strengths are demonstrated in
Fig.\ \ref{fig:hg194_pair}.  The top portion shows the pairing gap of
protons. The Krieger et al.\ strength ($G_{\rm p}$=0.192 MeV) gives a
rather large pairing gap ($\Delta_{\rm p} < 1.8$ MeV), while our
pairing strength 
($0.145 < G_{\rm p} < 0.168$ (MeV)) produces a reasonable size gap
($\Delta_{\rm p} < 1.2$ MeV) for $0.2 < \delta < 0.6$.
On the other hand, the pairing gap of neutrons does not differ very
much between the two calculations and has a reasonable size as shown
in the middle portion of Fig.\ \ref{fig:hg194_pair}.  This is a direct
consequence of the fact that the Krieger et al.\ strength ($G_{\rm
  n}=0.132$ MeV) and our strength ($0.114 < G_{\rm n} < 0.126$ (MeV))
do not differ very much.

The PES curves are shown in the bottom portion of Fig.\
\ref{fig:hg194_pair}.  One can see that the reduction of the proton's
pairing force strength raises the energy by 3 MeV at the barrier and
by 1 MeV at the SD minimum while leaving the energy almost unchanged
at the oblate minimum.
Consequently, Krieger et al.\ obtained $E^{\ast}=5.0$ MeV, which is
lower than our value by 1.3 MeV, and the barrier height
(defined in the figure) $E_{\rm B}=1.8$ MeV, which is lower than our
value by 1.7 MeV.

Note that the half-life of the SD band head due to decay into the
normal-deformaion (ND) well is longer for higher barriers.  We have
estimated the half-life using WKB approximation ($f_B = 10$ in Eq.\
(\ref{eq:fb})).
The resulting half-life is $6 \times 10^{-17}$ sec for the Krieger et al.\
pairing force strengths while it is $1 \times 10^{-13}$ sec for our strengths.
The difference amounts to a factor of about $10^{3}$.

In Fig.\ \ref{fig:nils}, we show the Nilsson diagram obtained by
self-consistent mean-field calculation for $^{194}$Hg using
SkM$^{\ast}$.  In the case of proton, one can see clearly a lacuna at
$Z=80$ in the vicinity of $\delta \sim 0.52$. On the other hand, there
is no clear lacuna for $N=114$ in the case of neutron. But the level
density near the Fermi surface is relatively small so that neighboring
isotopes have an SD. This is consistent with the fact that the proton
pairing gap sharply decreases, while the neutron pairing gap does
rather smoothly at $\delta \sim 0.52$, as we showed in Fig.\
\ref{fig:hg194_pair}.

The analysis of the SD state in $^{194}$Hg indicates the
appropriateness of our pairing force strengths combined with the
SkM$^{\ast}$, SkP, and SkI3 forces for the HF potential.
%

\section{Fission Isomers} \label{sect:fi}

We have also applied the HF+BCS method to fission isomers. Among several
tens of actinide nuclei having fission isomers, $^{236}$U and
$^{238}$U are the only ones whose fission isomers were observed to
decay electromagnetically into ND states\cite{TOI}. The energy of the
gamma-ray tells us the excitation energy of the fission isomer, while
the measured half-life gives information on the barrier between the SD
and the ND wells.

We estimate the half-lives as
\begin{equation} \label{eq:halflife}
  \label{tau}
  \tau_{1/2}=\frac{\ln 2}{nP},
\end{equation}
where the half-life $\tau_{1/2}$ is inversely proportional to two
quantities. One is the quantity, $n$, the number of assaults
in the SD well, which may be estimated as
\begin{equation}
  n=\frac{\omega_{\mbox{$\beta$-vib}}}{2\pi},
\end{equation}
where $\omega_{\mbox{$\beta$-vib}}$ is the angular frequency of the
$\beta$-vibration in the SD well.  We assume that
$\omega_{\mbox{$\beta$-vib}}$ scales as $A^{-1/6}$ and that
$\hbar\omega_{\mbox{$\beta$-vib}}=1$ MeV for $A=150$.  The other
quantity, $P$, is the penetration probability through the barrier into
the ND well, which is related to the action $S$ in the WKB
approximation as
\begin{eqnarray}
    P &=& \exp(-S),                               \label{eq:p} \\
    S &=& \frac{2}{\hbar} \int_{\beta_1}^{\beta_2}
    \sqrt{2B(\beta)(V(\beta)-E)} d\beta .         \label{eq:action}
\end{eqnarray}
On the right hand side \ of Eq.~(\ref{eq:action}),
$\beta_2$ is the deformation of the SD isomer while $\beta_1$ is
the deformation in the ND well where the total energy is the same
as that of the SD isomer, i.e., $V(\beta_1)=V(\beta_2)=E$.
We relate $\beta$ to $\delta$ defined by
Eq.~(\ref{eq:delta}) as
$\beta=(16 \pi / 45)^{1/2} \delta \simeq 1.06 \delta$.
The action $S$ takes on large values when the barrier is large and broad
and/or when the collective mass $B(\beta)$ is large.

The collective mass for small-amplitude irrotational hydro-dynamical
motions is expressed as\cite{RS80}
\begin{equation} \label{eq:b}
  B_{\lambda} = \frac{\rho R_0^5}{\lambda} =
  \frac{3 A m_N R_0^2}{4\lambda\pi},
\end{equation}
where $\lambda(=2)$ is the multipolarity of the motion, $\rho$ the
nucleon density inside the nucleus $(=3Am_N/4\pi R_0^3)$, $R_0$ the
nuclear radius ($= 1.2A^{1/3}$ fm), $A$ the mass number, and $m_N$ the
bare nucleon mass.
We express the collective mass in units of $B_2$, neglecting the
dependence on deformation:
\begin{equation} \label{eq:fb}
  B(\beta) = B_2 f_B.
\end{equation}
The empirical enhancement factor $f_B$ varies between 10 and 40 when it
reproduces the quadrupole vibrational levels of
spherical nuclei\cite{TH56}.


Table~\ref{tab:uran} gives the experimental and theoretical values of
the excitation energies and the partial half-lives of the fission
isomers.
The first line gives the experimental values, while the second
line shows the results calculated by Chinn et al.\ \cite{CBG92}.
They employed the Hartree-Fock-Bogoliubov
(HFB) method with the Gogny D1 force\cite{DG80}.
They calculated the half-lives by reducing the Hill-Wheeler
equation for the generator coordinate method (GCM) to a Schr\"odinger
equation with the Gaussian overlap approximation.
%
The fifth and the sixth lines give the results obtained by Krieger et
al.\ \cite{KBF96}, who employed the HF+BCS method with the
SkM$^{\ast}$ force and the pairing force strengths fitted so as to
reproduce the experimental pairing gaps.
They estimated the half-lives with the GCM without further
approximations.
Lines 7-10 are the results of our calculations using the SkM$^{\ast}$
force.
We determine the pairing force strengths in the same manner as
we did for $^{194}$Hg.

Chinn et al.\ reproduced the excitation energy of $^{238}$U rather
well (the error is +11\%).
In Ref.\ \cite{KBF94}, Krieger et al.\ obtained a better agreement with
the experimental value (the error is -11\%) by taking into account the
correlations in both the ground and SD states with the GCM method than
without the correlation (the error is +17\%).
Recently, they revised the calculations \cite{KBF96} with
slightly weaker pairing strengths so as to reproduce the experimental
pairing gaps ($G_{\rm n}=0.1067 \rightarrow 0.0955$ [MeV],
$G_{\rm p}=0.1621 \rightarrow 0.1505$ [MeV]).
As a result, they obtained improved results.
For $^{238}$U, the error is +6\% for HF+BCS, +2\% with GCM.
For $^{236}$U, the error is +5\% for HF+BCS, +1\% with GCM.  This also
shows that the excitation energy of the very deformed isomers is very
sensitive to the strengths of pairing force.


Figs.~\ref{fig:u236} and \ref{fig:u238} show the PES curves of
$^{236, 238}$U, respectively, which we have calculated employing the
SkM$^{\ast}$, SIII, and SkP forces.  The excitation energies for
$^{236}$U and $^{238}$U are overestimated by 1.8 MeV and 0.7 MeV,
respectively, with the SkM$^{\ast}$ force.  Using SIII or SkP makes
the agreement worse.

The partial half-lives were not reproduced very accurately either by
Chinn et al.\ \cite{CBG92} or Krieger et al.\ \cite{KBF96}.
As for our calculations, we have to determine the enhancement
factor $f_B$ before estimating the half-lives. Conversely,
when the factor  $f_B$ is adjusted to reproduce the experimental
half-lives, the resulting values are $f_B=13.7$ for $^{236}$U and
$f_B=10.4$ for $^{238}$U.

Our calculations of the half-lives are crude because of our use
of the WKB approximation.
There are many factors which can strongly influence the calculation of
the half-lives.
Our calculation takes into account the landscape of the potential energy
curve and the collective mass.
It does not take into account the triaxial deformation,
the level density in the ND well and the angular momentum projection effect.


\section{Systematics of the non-rotating SD} \label{sect:syshf}

From the analysis in the last two sections, SkM$^{\ast}$ seems the
best Skyrme force to describe non-rotating SD states. With this force,
we have explored a wide area of the nuclear chart ranging from $Z=20$
to $Z=82$ for the SD states at zero spin in even-even nuclei. To
compare the energy of zero spin states with experiments, one should
perform the angular momentum projection. In this paper, we approximate
the zero-spin states by non-rotating states.


An earlier microscopic attempt to explore SD at zero spin was made by
Bonche et al.\ \cite{BKQ89} for the Os-Pt-Hg region using the HF+BCS
with the SIII force. Later, Krieger et al.\ \cite{KBW92} changed
the force to SkM$^{\ast}$ and performed an extensive calculation covering
from $_{62}$Sm$_{126}$ to $_{92}$U$_{146}$ to obtain SD minima for 148
nuclei.

We employed the same Skyrme force as Krieger et al.\ used.
However, the results of our calculations are significantly different
as we will show at the end of this section.
The difference originates in their proton pairing-force strength,
which seems too strong as demonstrated clearly in sect.~\ref{sect:hg}
for $^{194}$Hg.

Our calculations do not cover very neutron-rich nuclei, unlike the
calculations by Krieger et al.,
because the pairing correlation among neutrons cannot be
correctly described within the HF+BCS scheme for these nuclei.
When the Fermi level approaches zero from below,
the continuum single-particle states are involved
strongly in the pairing correlation.
This coupling cannot be treated in the HF+BCS scheme,
which relies on an assumption that the pair-scattering matrix
elements are independent of the single-particle wave functions
(e.g., constant as in the seniority force).
For the correct description of the coupling, one has to switch
from the HF+BCS to the HFB scheme\cite{DFT84}.

In order to search for the SD solutions, we took the following steps:

i) We prepared an initial wave function by either using the solution
   for a neighboring nucleus or taking the eigenstate of the Nilsson
   potential of appropriate deformation.

ii) If the quadrupole deformation parameter $\delta$ of the initial
   wave function was smaller than 0.6, we exerted an external potential
   proportional to $\hat{Q}_z$ on the initial wave function until
   $\delta$ exceeded 0.6.  Then, we switched off the external potential.

iii) We let the wave function evolve in imaginary time \cite{DFK80} by
   itself. If it converged to a local minimum with $\delta > 0.35$, we
   regarded that the nucleus has an SD isomeric state. If $\delta$
   became less than 0.35 in the course of the self evolution, we
   concluded that the nucleus does not have an SD state.

iv) If the nucleus had an SD minimum, we calculated the potential
  energy curve for $0 \leq \delta \leq 0.6$
  ($-0.3 \leq \delta \leq 0.6$ for $Z$=80) by imposing a constraint
  on $\hat{Q}_z$. We also imposed $\langle x^2 -y^2 \rangle_{\rm mass}=0$
  for the sake of a fast convergence.
  This step required more than ten times as long
  computation time as the previous three steps.
  It was necessary, however, to calculate the excitation energy
  and the barrier height, which are necessary to estimate the half-life
  of the isomer.

Following the above prescription, we explored 642 even-even nuclei
with $20 \le Z \le 82$ and found SD minima in 118 nuclei.
%
%
In Table~\ref{tab:sd}, we give the deformation $\delta$, the
excitation energy $E^{\ast}$, and the barrier height $E_{\rm B}$ of
the thus-obtained large-deformation (i.e., $\delta > 0.35$) solutions
of 118 nuclei.


These results are presented again graphically in Fig.\ \ref{fig:def_e}.
The top portion shows the size of the quadrupole deformation $\delta$ in
terms of the diameter of the circles.

The middle portion gives the excitation energy measured from the
ground state. The diameter of the circles is inversely proportional to
the excitation energy. Circles are solid (open) when the excitation
energy is less (greater) than 8 MeV: SD states designated by solid
circles are easier to observe experimentally than those symbolized by
open circles.  When the SD state is the ground state, the circle is
not drawn.

The bottom portion shows the barrier height between the SD minimum and
the ground state in terms of the diameter of circles.  The circles are
solid (open) when the barrier height is greater (less) than 0.5 MeV:
SD states indicated by open circles are unlikely to have long
half-lives.
When the SD state is the ground state, the circle is not drawn.

The SD states obtained from our calculations can be classified
into nine groups.

\renewcommand{\labelitemi}{ }
\begin{itemize}

\item 1. $34 \le Z \le 42, 36 \le N \le 42$ (12 nuclei)\\
  These nuclei around $^{76}_{38}$Sr$_{38}$ are superdeformed in the
  ground state.  The axis ratio is about 3:2 ($\delta \sim 0.38$).
  They have no other prolate minima in the range $0 < \delta < 0.6$.

\item 2. $44 \le Z \le 48, 44 \le N \le 50$ (10 nuclei)\\
  These nuclei around $^{92}_{46}$Pd$_{46}$ have the largest
  deformations ($\delta \geq 0.6$) among the nuclei which we
  calculated.  These SD states are, however, difficult to detect
  because the excitation energies are rather high ($> 10$ MeV) while
  the barrier heights are rather low ($< 1$ MeV).

\item 3. $58 \le Z \le 60, 60 \le N \le 66$ (6 nuclei)\\
  These nuclei around $^{122}_{\phantom{0}58}$Ce$_{64}$ have rather
  large deformations ($\delta \sim 0.35$) in the ground state.

\item 4. $68 \leq Z \leq 82, 80 \leq N \leq 102$ (29 nuclei)\\
  Many nuclei have SD minima in this region near the proton-drip line.
  The deformations are in the range $0.41 \le \delta \le 0.63$.
  Some of them have low $E^{\ast}$ and high $E_{\rm B}$.
  For example, $^{172,174,176}$Hg have
  $E^{\ast}$=5.2, 4.5, and 4.1 MeV and $E_{\rm B}$=2.9, 2.0, and 1.2
  MeV, respectively.

\item 5. $74 \leq Z \leq 82, 106 \leq N \leq 134$ (48 nuclei)\\
  This group corresponds to the $A \sim 190$ region where many high-spin SD
  rotational bands have been observed experimentally. The deformations
  are in the range $0.46 \le \delta \le 0.55$.


  In Fig.\ \ref{fig:pb} we show how the landscape of the PES curve
  changes along the $Z=82$ isotope chain, which runs through this and
  the previous groups.
  One can see that, approaching to the spherical
  magic of $N=126$, the barrier height becomes higher while the
  excitation energy also increases.
  Indeed, $^{208}_{\phantom{2}82}$Pb$_{126}$ has the largest
  $E^{\ast}$ (23 MeV) and $E_{\rm B}$ (6.5 MeV) in this group.
  The former effect makes the half-lives of the SD states in
  $N \sim 126$ nuclei very long according to
  Eqs.~(\ref{eq:halflife})-(\ref{eq:action}).
  However, the latter effect makes those states difficult to detect
  experimentally.

  Judging from the location of each nucleus in ($E^{\ast}$, $E_{\rm
    B}$) plane, we think that the most promising nuclei for
  experimental observations in this group are $^{188,190,192}$Hg,
  which have $E^{\ast}$=3.7, 4.1, and 4.9 MeV and $E_{\rm B}$=1.1,
  2.4, and 3.6 MeV, respectively. On the other hand, $^{194}$Hg,
  discussed in sect.~\ref{sect:hg}, is less promising than these three
  isotopes because it has a higher $E^{\ast}$ but a little smaller
  $E_{\rm B}$ than $^{192}$Hg has.

\item 6. $38 \le Z \le 40, 60 \le N \le 68$ (8 nuclei)\\
  Nuclei near $^{104}_{\phantom{1}40}$Zr$_{64}$ have rather large
  deformations ($\delta \sim 0.35$) in the ground state.

\item 7. $48 \le Z \le 50, N=80$ (2 nuclei)\\
  $^{128}_{\phantom{1}48}$Cd$_{80}$ and
  $^{130}_{\phantom{1}50}$Sn$_{80}$ have SD minima at $\delta \sim
  0.56$.
  The quantum fluctuation of shape is expected to wash out
  these minima because the excitation energy is very high ($> 20$ MeV)
  while the barrier is very low ($< 130$ keV).

\item 8. $Z=58, N=80$ (1 nucleus)\\
  $^{138}_{\phantom{1}58}$Ce$_{80}$ has a SD minimum at $\delta =
  0.44$, which seems too shallow (the barrier height is 120 keV) to
  confine the collective wave function.

\item 9. $Z=64, 90 \le N \le 92$ (2 nuclei)\\
  $^{154}_{\phantom{1}64}$Gd$_{90}$ and
  $^{156}_{\phantom{1}64}$Gd$_{92}$ have SD minima with very low
  barriers ($\sim 80$ keV).
  The PES of a neighboring isotope
  $^{152}$Gd is plotted in Fig.\ \ref{fig:gd}, where the SD
  minimum existing in $^{154}$Gd has become a shoulder.
  The behavior of the pairing gaps shown in the figure
  suggests that the SD shell gaps occur at somewhat
  different deformations between protons and neutrons.

  These nuclei are in the $A \sim 150$ region where high-spin SD
  rotational bands have been discovered in many nuclei. However, only
  two nuclei have the SD minimum at zero spin in contrast to the
  situation for the $A \sim 190$ region.  This is in agreement with
  the fact that SD rotational bands are found as low as $I = 20 (10)
  \hbar$ in this ($A \sim 190$) mass region.

   Incidentally, we have not found any zero-spin SD states in nuclei
   belonging to the $A \sim 130$ mass region of high-spin SD
   rotational bands.

\end{itemize}
\renewcommand{\labelitemi}{$\bullet$}

In Figs.~\ref{fig:ex} and \ref{fig:eb}, we compare $E^{\ast}$ and
$E_{\rm B}$ of Krieger et al.\ with our results.  One can see that the
excitation energies of the SD states are not very different while our
barrier heights are about twice as high as the values of Krieger et
al.  This change is brought about by our weaker pairing force
strengths.

\section{Comparison with the results of the Nilsson-Strutinsky method}
\label{sect:ns}

The Nilsson-Strutinsky(NS) method is a convenient and
well-established method to treat nuclear deformations.
Because NS computation is much simpler than self-consistent-field
calculations, it seems worth doing another systematic survey of SD
isomers using the NS method in order to compare the results
with those of the Skyrme-HF method for the same nuclei.
Our survey using NS method provides a useful overview of the situation
outside the region investigated in the last section.

We have utilized a program for the standard Nilsson-Strutinsky
calculation\cite{NTS69} provided by Y.~R.~Shimizu\cite{Shi97},
which takes into account two axially symmetric deformations, i.e., the
quadrupole deformation $\epsilon_2$ and the hexadecapole deformation
$\epsilon_4$.  For each value of $\epsilon_2$, the value of
$\epsilon_4$ is optimized so as to minimize the total energy.

The standard values given in Table~1 of Ref.\ \cite{BR85} are used for
the parameters $\kappa_N$ and $\mu_N$ of the Nilsson potential.  The
pairing correlation is active for single-particle levels within $\pm
1.2 \hbar \omega$ from the Fermi level, while the strengths of the
pairing force are determined such that the smoothed pairing gap
becomes $\bar{\Delta}$ = $ 13 A^{-1/2}$ MeV.  The parameters of the
macroscopic part\cite{MS67}
are $a_{\rm s}=17.9439$ MeV, $\kappa_{\rm s}=1.7826$,
and $R_{\rm c}=1.2249 A^{1/3}$ fm.
See Ref.\ \cite{BRA91} for calculational details.

With this model one can calculate the entire region of the nuclear
chart, i.e., from the proton drip line to the neutron drip line: The
model does not suffer from the problem of neutron pairing in
neutron-rich nuclei explained in sect.~\ref{sect:syshf} because the
Nilsson potential does not have a continuum spectrum.  Concerning the
expected enhancement of the pairing due to the coupling with the
continuum states, however, the present model simply neglects its
influences.

The calculation for 2000 even-even nuclei can be completed in a few
hours with an ordinary personal computer owing to the simpleness of
the NS method itself and also to the specialization of the code to
non-rotating axially symmetric states.

Fig.\ \ref{fig:def_n} displays the resulting deformations $\epsilon_2$,
excitation energies $E^{\ast}$, and barrier heights $E_{\rm B}$
of the SD minima at $\epsilon_2 > 0.35$.  Let us discuss the results
according to the grouping employed in sect.~\ref{sect:syshf}.

\renewcommand{\labelitemi}{ }
\begin{itemize}

\item 1. $38 \le Z \le 40, 36 \le N \le 38$\\
  The area of this island of zero-spin SD states is much smaller than
  in the Skyrme-HF results.

\item 2. $48 \le Z \le 50, 44 \le N \le 48$\\
  The number of nuclei is reduced from 10 to 4.

\item 3. $58 \le Z \le 60, 60 \le N \le 66$\\
  None of the 6 nuclei were found to display SD isomers in the NS model.

\item 4. $70 \leq Z \leq 80, 80 \leq N \leq 102$\\
  The number of nuclei is almost unchanged (thirty), but the
  number of neutron deficient (rich) nuclei is decreased (increased).
  The appearance of SD around Pb-Hg-Pt isotopes is quite similar in
  the Skyrme-HF and the NS methods.  However, the details are
  different. For example, the PES of the Skyrme-HF has a smaller
  $E^{\ast}$ and a larger $E_{\rm B}$ than that of the NS model for
  the Pb isotopes.

\item 5. $Z+ N > \sim 200$\\
  One can see that the group No.~5 found in the Skyrme-HF is a part of
  a huge area extending to $Z>82$ and/or $N>126$.  In the actinide
  region there exist fission isomers for almost all the nuclei except
  in a rectangle-like area $Z\ge 102$ and $N \le 186$, where
  large deformation minima do not exist and the nucleus goes into
  fission directly from the ground state.

\item 6. $38 \le Z \le 40, 60 \le N \le 68$\\
  None of 8 nuclei in this region show SD isomers.

\item 7,8. $36 \le Z \le 62, 78 \le N \le 82$\\
  This region extending vertically includes small islands of Nos.~7 and 8
  listed in sect.~\ref{sect:syshf}.

\item 9. $Z=68, 86 \le N \le 88$\\
  Compared with the island No.~9 of the Skyrme-HF results, SD isomers are
  found at $Z$ and $N$ values differing by +4 and $-4$, respectively.

\item 10. $46 \le Z \le 54, 98 \le N \le 102$\\
  The nine nuclei above the neutron-drip line have zero-spin SD states.
  The deformations are in the range $0.63 \le \delta \le 0.64$.

\item 11. $54 \le Z \le 66, 118 \le N \le 130$\\
  Each of twenty-seven nuclei above the neutron-drip line has
  a large-deformation ($0.36 \le \delta \le 0.41$) solution
  mostly in the ground state.

\end{itemize}
\renewcommand{\labelitemi}{$\bullet$}

\section{Conclusions} \label{sect:sum}

In this paper we studied the super-deformed states at zero spin
using the HF+BCS method with the Skyrme interactions.

In sect.~\ref{sect:hg}, we compared nine parameter sets of the
Skyrme force in their ability to reproduce the SD bandhead of
$^{194}$Hg.  The best agreements of the excitation energy with the
experimental value are obtained for the SkM$^{\ast}$, SkP, and SkI3
interactions.
The contour of the PES curve, especially the barrier
height, is found to be very sensitive to the pairing force strengths.

In sect.~\ref{sect:fi}, we presented our studies of the fission
isomers in  $^{236,238}$U.
The agreement with experimental excitation energies
is better for the SkM$^{\ast}$ force than for the SkP and SIII
forces.  The SkM$^{\ast}$ seems to be the best one among the
available parameter sets of the standard Skyrme interaction for
the treatment of large deformations.

In sect.~\ref{sect:syshf}, we presented our studies of the systematics
of large-deformation ($\delta > 0.35$) states of even-even nuclei
using the SkM$^{\ast}$ force.  We searched 642 even-even nuclei with
$20 \leq Z \leq 82$ and found 118 SD minima.  We analyzed the
systematics of the deformation, excitation energy, and barrier height
of these SD minima.

The difference between our calculation and an earlier systematic study
by Krieger et al.\ using the same Skyrme force lies in the strengths
of the pairing force.
We determine the strengths such that the empirical
formula $\bar{\Delta}=12 A^{-1/2}$ MeV is reproduced for the
averaged single-particle level density using the Thomas-Fermi
approach.
Our strengths determined in this manner are weaker than
those adopted by Krieger et al. As a consequence, the barrier heights
are doubled while the excitation energies are not changed so much.

In sect.~\ref{sect:ns}, we also used the Nilsson-Strutinsky
method to do another systematic comparison with the results of
our Skyrme-HF calculation.


The results of our calculations presented in Table~\ref{tab:sd}
are available electronically through the Internet at
{\bf http://nt2.c.u-tokyo.ac.jp/hf/sdzs/}.

The authors thank Dr.~P.~Bonche, Dr.~H.~Flocard, and Dr.~P.-H.~Heenen
for providing the HF+BCS code {\sl EV8}.  They are also grateful to
Dr.~Y.R.~Shimizu for providing the Nilsson-Strutinsky code and
commenting on the corresponding part of this paper.  We thank
Dr.~P.-H. Heenen, Dr.~M.S. Weiss and Dr.~S.J.Krieger for useful
discussions.  A part of this work was financially supported by RCNP,
Osaka University, as RCNP Computational Nuclear Physics Project
(No.~95-B-01).  The remaining part of the calculations were performed
with a super computer VPP500 at RIKEN (Research Institute for Physical
and Chemical Research, Japan).


\newpage

\noindent{\bf FIGURE CAPTIONS}

\newcounter{figno}

\begin{list}
  {Fig. \arabic{figno}. }{\usecounter{figno}
    \setlength{\labelwidth}{1.3cm} \setlength{\labelsep}{0.5mm}
    \setlength{\leftmargin}{8.5mm} \setlength{\rightmargin}{0mm}
    \setlength{\listparindent}{0mm} \setlength{\parsep}{0mm}
    \setlength{\itemsep}{0.5cm} \setlength{\topsep}{0.5cm} }

  \baselineskip=0.821cm

\item \label{fig:hg194} 
  Potential energy curves of $^{194}$Hg for nine parameter sets of the
  Skyrme force. The abscissa represents the axially symmetric
  quadrupole deformation parameter $\delta$ while the ordinate denotes
  the energy measured from the sphericity. In parentheses are the
  excitation energies (in MeV) of the superdeformed minima measured
  from the ground-state minima at $\delta \sim -0.1$.

\item \label{fig:hg194_pair} 
  Proton (top) and neutron (middle) pairing gaps and the potential
  energy curve (bottom) of $^{194}$Hg versus the deformation parameter
  $\delta$ calculated with the SkM$^{\ast}$ force and two sets of pairing
  force strengths.  The solid (dash) curves are calculated with our
  (Krieger's) strengths. We determine the strengths such that the
  empirical formula $\bar{\Delta}=12 A^{-1/2}$ MeV is satisfied
  for Thomas-Fermi level density.

\item \label{fig:nils}
  Neutron(left) and proton(right) single-particle energies versus quadrupole
  deformation parameter $\delta$ for $^{194}$Hg obtained by the
  HF+BCS using SkM$^{\ast}$. Positive-parity states are drawn by solid
  lines and negative-parity states by dashed lines.

\item  \label{fig:u236}  
  Potential energy curves of $^{236}$U versus the deformation
  parameter $\delta$ calculated with the SkM$^{\ast}$, SIII, and
  SkP forces.  The ordinate denotes the energy measured from the
  normal-deformation minimum. The experimental point is also shown
  with a horizontal error bar.

\item  \label{fig:u238}  
  Same as in Fig.\ \ref{fig:u236} but for $^{238}$U.

\item  \label{fig:def_e}  
  Properties of the superdeformed (i.e., $\delta > 0.35$) minima
  calculated with the HF+BCS using SkM$^{\ast}$ force.  The top portion
  displays the quadrupole deformation parameter $\delta$.  Nuclei
  having SD minima are designated with circles whose diameter is
  proportional to $\delta$ of the SD state. Open (solid) circles are
  used when the SD state is the ground state (an excited state).  In
  the middle portion, the diameter of the circles is inversely
  proportional to the excitation energy of the SD minimum.  In the
  bottom portion, the diameter is proportional to the well depth of
  the SD minimum.  Except in the top portion, a circle is not drawn
  when the SD minimum is the ground state.  The grid indices the
  locations of the magic numbers for spherical shape. The solid
  staircase-like lines represent the two-proton and the two-nucleon
  drip lines taken from Ref.\ \cite{TTO96}.  Our calculations have been
  done only for the nuclei between the two-proton drip line and the
  dashed staircase-like line.

\item  \label{fig:pb}  
  Potential energy curves of Pb isotopes versus the
  deformation parameter $\delta$. The ordinate denotes the energy measured
  from the sphericity. The left, the middle, and the right portions include
  the curves for $N$=94-102, $N$=104-124, and $N$=126-134, respectively.

\item  \label{fig:gd}  
  Proton (top) and neutron (middle) pairing gaps and the potential
  energy curve (bottom) of $^{152}$Gd versus the deformation parameter
  $\delta$ calculated with the SkM$^{\ast}$ force.

\item \label{fig:ex} 
  Comparison of the excitation energy $E^{\ast}$ of the zero-spin SD states
  between Krieger's and our results for $Z=76-82$ even-even isotope chains.

\item \label{fig:eb} 
  Same as in Fig.\ \ref{fig:ex} but for
  the barrier heights between the SD and ND wells.

\item  \label{fig:def_n} 
  Same as in Fig.\ \ref{fig:def_e} but with the
  Nilsson-Strutinsky method.

\end{list}


\newpage

\noindent {\bf TABLES}

\begin{table}[htbp]
  \begin{center}
    \caption{Comparison of the excitation energy of the SD state at
      zero-spin in $^{194}$Hg between the
      (extrapolated) experimental value and
      the predictions of earlier theoretical works.
     }
    \vspace*{5mm}
    \leavevmode
    \begin{tabular}{ll}
      \hline
                                             & $E^{\ast}$ (MeV) \\
      \hline
      experiment\cite{Kh96}                  & 6.017  \\
      Woods-Saxon-Strutinsky\cite{Sa91}      & 4.6    \\
      Woods-Saxon-Strutinsky\cite{Ch89,Kh96} & 4.9    \\
      HF+BCS with SkM$^{\ast}$\cite{KBW92}   & 5.0    \\
      HFB with Gogny D1      \cite{De89}     & 6.9    \\
      Nilsson-Strutinsky     \cite{Ri90}     & 7.5    \\
      \hline
    \end{tabular}
    \label{tab:hg194}
  \end{center}
\end{table}


\begin{table}[htbp]
  \begin{center}
    \caption{Excitation energies $E^{\ast}$ and half-lives $\tau_{1/2}$ of
      the fission isomers in $^{236, 238}$U.
      The experimental values are compared with the theoretical results
      by Chinn et al.\ and Krieger et al.\ as well as with our results
      using the SkM$^{\ast}$ force.
      As for the half-lives, our calculation depends on the
      enhancement factor for the collective mass $f_{\rm B}$.}
    \vspace*{5mm}
    \leavevmode
    \begin{tabular}{lllll}
      \hline
      & \multicolumn{2}{c}{$E^{\ast}$ (MeV)}
      & \multicolumn{2}{c}{$\tau_{1/2}$ (sec)}                     \\
      \hline
      & $^{236}$U & $^{238}$U & $^{236}$U  & $^{238}$U             \\
      \hline
      experiment \cite{TOI} &
      2.750 & 2.557 & 1.20$\times 10^{-7}$ & 2.98 $\times 10^{-7}$ \\
      Chinn et al.\ \cite{CBG92} &
      $-$   & 2.828 &  $-$                 & 8.5 $\times 10^{-5}$  \\
      Krieger et al.\ \cite{KBF94} w/o GCM &
      $-$   & 3.0   &  $-$                 & $-$                   \\
      Krieger et al.\ \cite{KBF94} with GCM &
      $-$   & 2.28  &  $-$                 & 0.05-20 $\times 10^{-9} $\\
      Krieger et al.\ \cite{KBF96} w/o GCM &
      2.9   & 2.7   &  $-$                 & $-$                   \\
      Krieger et al.\ \cite{KBF96} with GCM &
      2.77  & 2.61  & $3.7 \times 10^{-8}$ & $8.5 \times 10^{-9} $ \\
      ours with $f_{\rm B}$=10 &
      4.5   &  3.3  & 1.3 $\times 10^{-9}$ & 1.6 $\times 10^{-7}$  \\
      ours with $f_{\rm B}$=20 &
      4.5   &  3.3  & 8.4 $\times 10^{-5}$ & 7.8 $\times 10^{-2} $ \\
      ours with $f_{\rm B}$=30 &
      4.5   &  3.3  & 4.1 $\times 10^{-1}$ & 1.8 $\times 10^{3} $  \\
      ours with $f_{\rm B}$=40 &
      4.5   &  3.3  & 5.3 $\times 10^{2}$  & 8.4 $\times 10^{6} $  \\
      \hline
    \end{tabular}
    \label{tab:uran}
  \end{center}
\end{table}


\begin{table}[htbp]
  \begin{center}
    \caption{Properties of SD minima in $20 \le Z \le 82$ even-even nuclei
       calculated with the HF+BCS method using the SkM$^{\ast}$ force and
       a seniority pairing force.
       The columns present the atomic number $Z$, the neutron number $N$,
       the deformation $\delta$, the excitation energy $E^{\ast}$, and
       the barrier height $E_{\rm B}$.
       A letter ``g'' in the first column indicates that the
       SD minimum is the ground state, while a chatacter ``$\ast$''
       means that it is an excited state having a rather low
       excitation energy ($<8$ MeV) and not a too shallow barrier height
       ($>0.5$ MeV).}
    \vspace*{5mm}
    \leavevmode
    \label{tab:sd}
  \end{center}
\end{table}


\twocolumn[Table 3:
\vspace*{5mm}
]
\begin{tabular}{crrrrr}
\hline & $Z$ & $N$ & $\delta$ & $E^{\ast}$ & $E_{\rm B}$ \\
       &     &     &          & (MeV)      & (MeV)       \\ \hline
 g & 36& 36& 0.38&      &     \\
 g & 36& 38& 0.38&      &     \\
 g & 36& 40& 0.36&      &     \\
   & 38& 34& 0.36&  0.23& 0.11\\
 g & 38& 36& 0.39&      &     \\
 g & 38& 38& 0.40&      &     \\
 g & 38& 40& 0.39&      &     \\
 g & 38& 42& 0.36&      &     \\
 g & 38& 60& 0.36&      &     \\
 g & 38& 62& 0.37&      &     \\
 g & 38& 64& 0.37&      &     \\
 g & 38& 66& 0.36&      &     \\
 g & 40& 38& 0.40&      &     \\
 g & 40& 40& 0.41&      &     \\
 g & 40& 62& 0.36&      &     \\
 g & 40& 64& 0.37&      &     \\
 g & 40& 66& 0.36&      &     \\
 g & 40& 68& 0.35&      &     \\
 g & 42& 38& 0.38&      &     \\
 g & 42& 40& 0.37&      &     \\
$*$& 44& 44& 0.62&  6.69& 0.53\\
   & 44& 46& 0.62&  9.91& 0.51\\
   & 44& 48& 0.61& 13.73& 0.23\\
   & 44& 50& 0.65& 16.28& 0.33\\
   & 46& 44& 0.63&  8.83& 0.86\\
   & 46& 46& 0.68& 11.88& 1.04\\
   & 46& 48& 0.64& 16.29& 0.33\\
   & 48& 44& 0.63& 11.70& 0.74\\
   & 48& 46& 0.66& 15.14& 0.64\\
   & 48& 48& 0.62& 19.62& 0.07\\
\hline \end{tabular} \newpage \begin{tabular}{crrrrr}
\hline & $Z$ & $N$ & $\delta$ & $E^{\ast}$ & $E_{\rm B}$ \\
       &     &     &          & (MeV)      & (MeV)       \\ \hline
   & 48& 80& 0.56& 21.80& 0.13\\
   & 50& 80& 0.56& 25.67& 0.05\\
 g & 58& 60& 0.35&      &     \\
 g & 58& 62& 0.36&      &     \\
 g & 58& 64& 0.36&      &     \\
   & 58& 80& 0.44&  9.31& 0.12\\
 g & 60& 62& 0.37&      &     \\
 g & 60& 64& 0.36&      &     \\
 g & 60& 66& 0.35&      &     \\
   & 64& 90& 0.61&  8.02& 0.08\\
   & 64& 92& 0.60&  9.53& 0.07\\
   & 68& 80& 0.51&  6.37& 0.00\\
   & 68& 84& 0.43&  7.13& 0.01\\
   & 70& 80& 0.53&  6.91& 0.02\\
   & 70& 82& 0.43& 10.86& 0.01\\
   & 70& 84& 0.43&  7.63& 0.12\\
   & 72& 80& 0.53&  7.56& 0.23\\
   & 72& 82& 0.43& 11.75& 0.66\\
   & 72& 84& 0.43&  8.47& 0.78\\
   & 72& 86& 0.41&  5.93& 0.35\\
   & 74& 84& 0.50&  9.96& 2.20\\
$*$& 74& 86& 0.48&  7.35& 1.44\\
   & 74& 88& 0.46&  5.84& 0.39\\
   & 74&114& 0.47&  6.25& 0.13\\
   & 74&116& 0.48&  6.66& 0.36\\
   & 74&118& 0.46&  6.14& 0.19\\
   & 76& 86& 0.52&  8.65& 3.27\\
$*$& 76& 88& 0.51&  6.76& 1.80\\
   & 76& 90& 0.47&  5.78& 0.46\\
   & 76&110& 0.47&  5.50& 0.12\\
\hline \end{tabular} \newpage
\twocolumn[Table 3 - continued
\vspace*{5mm}
]
\begin{tabular}{crrrrr}
\hline & $Z$ & $N$ & $\delta$ & $E^{\ast}$ & $E_{\rm B}$ \\
       &     &     &          & (MeV)      & (MeV)       \\ \hline
$*$& 76&112& 0.49&  5.93& 0.67\\
$*$& 76&114& 0.49&  6.47& 1.30\\
$*$& 76&116& 0.49&  7.18& 1.80\\
$*$& 76&118& 0.50&  7.23& 1.88\\
   & 76&120& 0.48&  8.23& 1.89\\
   & 76&122& 0.47& 10.19& 2.16\\
$*$& 78& 88& 0.55&  7.70& 3.80\\
$*$& 78& 90& 0.55&  5.95& 2.21\\
$*$& 78& 92& 0.54&  5.08& 0.86\\
   & 78& 94& 0.52&  4.57& 0.10\\
   & 78&108& 0.47&  3.47& 0.10\\
$*$& 78&110& 0.49&  3.76& 0.75\\
$*$& 78&112& 0.51&  4.21& 1.55\\
$*$& 78&114& 0.51&  4.85& 2.32\\
$*$& 78&116& 0.51&  5.89& 3.60\\
$*$& 78&118& 0.51&  7.06& 3.88\\
   & 78&120& 0.50&  9.21& 3.77\\
   & 78&122& 0.49& 12.13& 3.79\\
   & 78&124& 0.49& 15.07& 4.10\\
   & 78&126& 0.48& 18.07& 4.42\\
$*$& 80& 92& 0.56&  5.20& 2.91\\
$*$& 80& 94& 0.56&  4.51& 1.96\\
$*$& 80& 96& 0.55&  4.12& 1.20\\
   & 80& 98& 0.53&  3.96& 0.48\\
   & 80&100& 0.48&  3.60& 0.06\\
   & 80&106& 0.50&  3.45& 0.14\\
$*$& 80&108& 0.52&  3.69& 1.06\\
$*$& 80&110& 0.52&  4.12& 2.38\\
$*$& 80&112& 0.52&  4.86& 3.62\\
$*$& 80&114& 0.52&  6.27& 3.59\\
\hline \end{tabular} \newpage \begin{tabular}{crrrrr}
\hline & $Z$ & $N$ & $\delta$ & $E^{\ast}$ & $E_{\rm B}$ \\
       &     &     &          & (MeV)      & (MeV)       \\ \hline
$*$& 80&116& 0.52&  7.36& 5.41\\
$*$& 80&118& 0.52&  7.60& 5.82\\
   & 80&120& 0.52& 11.79& 5.73\\
   & 80&122& 0.51& 14.13& 5.79\\
   & 80&124& 0.51& 16.94& 5.89\\
   & 80&126& 0.50& 19.90& 6.14\\
   & 80&128& 0.51& 16.75& 6.00\\
   & 80&130& 0.51& 13.70& 6.06\\
$*$& 82& 94& 0.59&  6.13& 2.73\\
$*$& 82& 96& 0.59&  5.28& 1.97\\
$*$& 82& 98& 0.60&  4.72& 1.18\\
   & 82&100& 0.62&  4.14& 0.42\\
   & 82&102& 0.63&  3.78& 0.27\\
   & 82&106& 0.55&  3.84& 0.05\\
$*$& 82&108& 0.55&  4.10& 0.94\\
$*$& 82&110& 0.55&  4.65& 2.36\\
$*$& 82&112& 0.54&  5.58& 3.76\\
$*$& 82&114& 0.54&  6.86& 4.79\\
   & 82&116& 0.54&  8.55& 5.59\\
   & 82&118& 0.54& 10.75& 6.20\\
   & 82&120& 0.53& 13.84& 6.14\\
   & 82&122& 0.52& 17.11& 6.06\\
   & 82&124& 0.50& 20.17& 6.32\\
   & 82&126& 0.50& 23.05& 6.51\\
   & 82&128& 0.49& 19.75& 6.39\\
   & 82&130& 0.50& 16.69& 6.28\\
   & 82&132& 0.51& 13.72& 6.28\\
   & 82&134& 0.52& 10.85& 6.34\\
   &   &   &     &      &     \\
   &   &   &     &      &     \\
\hline
\end{tabular}

\onecolumn


\begin{thebibliography}{99}  
\setlength{\itemsep}{0cm}
\setlength{\parsep}{0cm}
\bibitem{TW86} P. J. Twin, B. M. Nyako, A. H. Nelson, J. Simpson,
  M. A. Bentley, H. W. Cranmer-Gordon, P. D. Forsyth, D. Howe,
  A. R. Mokhtar, J. D. Morrison, J. F. Sharpey-Schafer, G. Sletten,
  Phys.  Rev. Lett. {\bf 57} (1986) 811.

\bibitem{Ba95} C. Baktash, D. M. Cullen, J. D. Garrett, C. J. Gross,
  N. R. Johnson, W. Nazarewicz, D. G. Sarantites, J. Simpson,
  T. R. Werner , Phys. Rev. Lett. {\bf 74} (1995) 1946.

\bibitem{NT88}P. J. Nolan and P. J. Twin, Annu. Rev. Nucl. Part. Sci.
  {\bf 38} (1988) 533.

\bibitem{Ab90} S. {\AA}berg, Nucl. Phys. {\bf A520} (1990) 35c.

\bibitem{JK91} R. V. F. Janssens and T. L. Khoo,
  Annu. Rev. Nucl. Part. Sci. {\bf 41} (1991) 321.

\bibitem{TTO96} N.\ Tajima, S.\ Takahara and N.\ Onishi, Nucl.\
  Phys. {\bf A603} (1996) 23.


%
\bibitem{BFH85}  
         P.~Bonche, H.~Flocard, P.-H.~Heenen, S.J.~Krieger and
         M.S.~Weiss, Nucl.\ Phys.\ {\bf A443} (1985) 39.

\bibitem{KBW92} S. J. Krieger, P. Bonche, M. S. Weiss, J. Meyer, H. Flocard,
    P.-H. Heenen, Nucl. Phys. {\bf A542} (1992) 43.

\bibitem{Kh96}
  T. L. Khoo, M. P. Carpenter, T. Lauritsen, D. Ackermann, I. Ahmad,
  D. J. Blumenthal, S. M. Fischer, R. V. F. Janssens, D. Nisius, E. F.
  Moore, A. Lopez-Martens, T. Dossing, R. Kruecken, S. J. Asztalos, J.
  A. Becker, L. Bernstein, R. M. Clark, M. A. Deleplanque,
  R. M. Diamond, P. Fallon, L. P. Farris, F. Hannachi, E. A. Henry, A.
  Korichi, I. Y. Lee, A. O. Macchiavelli, F. S. Stephens ,
  Phys. Rev. Lett. {\bf 76} (1996) 1583.

\bibitem{Sa91}
         W. Satula, S. Cwiok, W. Nazarewicz, R. Wyss, A. Johnson,
         Nucl. Phys. {\bf A529} (1991) 289.

\bibitem{Ch89}
         R. Chasman, Phys. Lett. {\bf B219} (1989) 227.

\bibitem{De89}
         J. P. Delaroche, M. Girod, J. Libert, I. Deloncle,
         Phys. Lett. {\bf B232} (1989) 145.

\bibitem{Ri90}
         M. A. Riley, D. M. Cullen, W. Nazarewicz, A. Alderson,
         I. Ali, P. Fallon, P. D. Forsyth, F. Hanna, S. M. Mullins,
         J. W. Roberts, J. F. Sharpey-Schafer, P. J. Twin, R. J. Poynter,
         R. Wadsworth, M. A. Bentley, A. M. Bruce, J. Simpson, G. Sletten,
         T. Bengtsson, R. Wyss , Nucl. Phys. {\bf A512} (1990) 178.
%

\bibitem{BFG75} 
         M. Beiner, H. Flocard, Nguyen van Giai and P. Quentin,
         Nucl.Phys. {\bf A238} (1975) 29.

\bibitem{APD92} 
         Y.~Aboussir, J.M.~Pearson, A.K.~Dutta and F.~Tondeur,
         Nucl.\ Phys.\ {\bf A549} (1992) 155;
         Atomic Data Nucl. Data Tables {\bf 61} (1995) 127.

\bibitem{BQB82} 
         J. Bartel, P. Quentin, M. Brack, C. Guet and
         H.-B. Hakansson, Nucl.Phys. {\bf A386} (1982) 79.

\bibitem{DFT84} 
         J.~Dobaczewski, H.~Flocard and J.~Treiner,
         Nucl.Phys. {\bf A422} (1984) 103.

\bibitem{GS81} 
         Nguyen van Giai and H. Sagawa,
         Phys.Lett {\bf B106} (1981) 379.

\bibitem{RF94} 
         P.-G.\ Reinhard and H.\ Flocard,
         Nucl.\ Phys.\ {\bf A584} (1995) 467.


\bibitem{TOI}
         R. B. Firestone, V. S. Shirley, S. Y. F. Chu,
         C. M. Baglin and J. Zipkin, {\em Table of Isotopes, John Wiley
         and Sons, New York} (1996).

\bibitem{RS80} 
         P.~Ring and P.~Schuck, The nuclear many-body problem
         (Springer, New York, 1980), sect.~1.4.

\bibitem{TH56}
        G. M. Temmer, N. P. Heydenberg, Phys. Rev. {\bf 104} (1956) 967.

\bibitem{CBG92}
        C. R. Chinn, J. -F.  Burger, D. Gogny and M. S. Weiss,
        Phys. Rev. {\bf C45} (1994) 1700.

\bibitem{DG80} 
         J.~Decharg\'{e} and D.~Gogny,
         Phys. Rev. {\bf C21} (1980) 1568.

\bibitem{KBF94} 
  S. J. Krieger, P. Bonche, H. Flocard, P. -H. Heenen,
  and M. S. Weiss, Nucl. Phys. {\bf A572} (1994) 384.

\bibitem{KBF96} 
          S.J.~Krieger, P.~Bonche, H.~Flocard, P.-H.~Heenen,
          R.~Mehrem and M.S.~Weiss,  Phys.\ Rev.\ {\bf C54} (1996) 2399.

\bibitem{BKQ89}
         P. Bonche, S. J. Krieger, P. Quentin, M. S. Weiss,
         J. Meyer, M. Meyer, N. Redon, H. Flocard and P. -H. Heenen,
         Nucl. Phys. {\bf A500} (1989) 308.

\bibitem{DFK80}
         K.T.R. Davis, H.Flocard, S.J. Krieger and M.S. Weiss,
         Nucl. Phys. {\bf A342} (1980) 111.

\bibitem{NTS69} 
         S.G.~Nilsson, C.F.~Tsang, A.~Sobiczewski, Z.~Szyma\'{n}ski,
         S.~Wycech, C.~Gustafson, I.~Lamm, P.~M\"{o}ller and B.~Nilsson,
         Nucl.Phys. {\bf A131} (1969) 1.

\bibitem{Shi97} 
         Y.R.~Shimizu, private communication.

\bibitem{BR85} 
         T. Bengtsson and I. Ragnarsson, Nucl. Phys. {\bf A436} (1985) 14.

\bibitem{MS67}
         W.D.~Myers and W.J.~Swiatecki, Ark.\ Phys.\ {\bf 36} (1967) 343.

\bibitem{BRA91} 
         T.~Bengtsson, I.~Ragnarsson and S.~{\AA}berg,
         in ``Computational Nuclear Physics 1'',
         ed. K.~Langanke, J.A.~Maruhn and S.E.~Koonin,
         (Springer-Verlag, Berlin, 1991) 51.



\end{thebibliography}
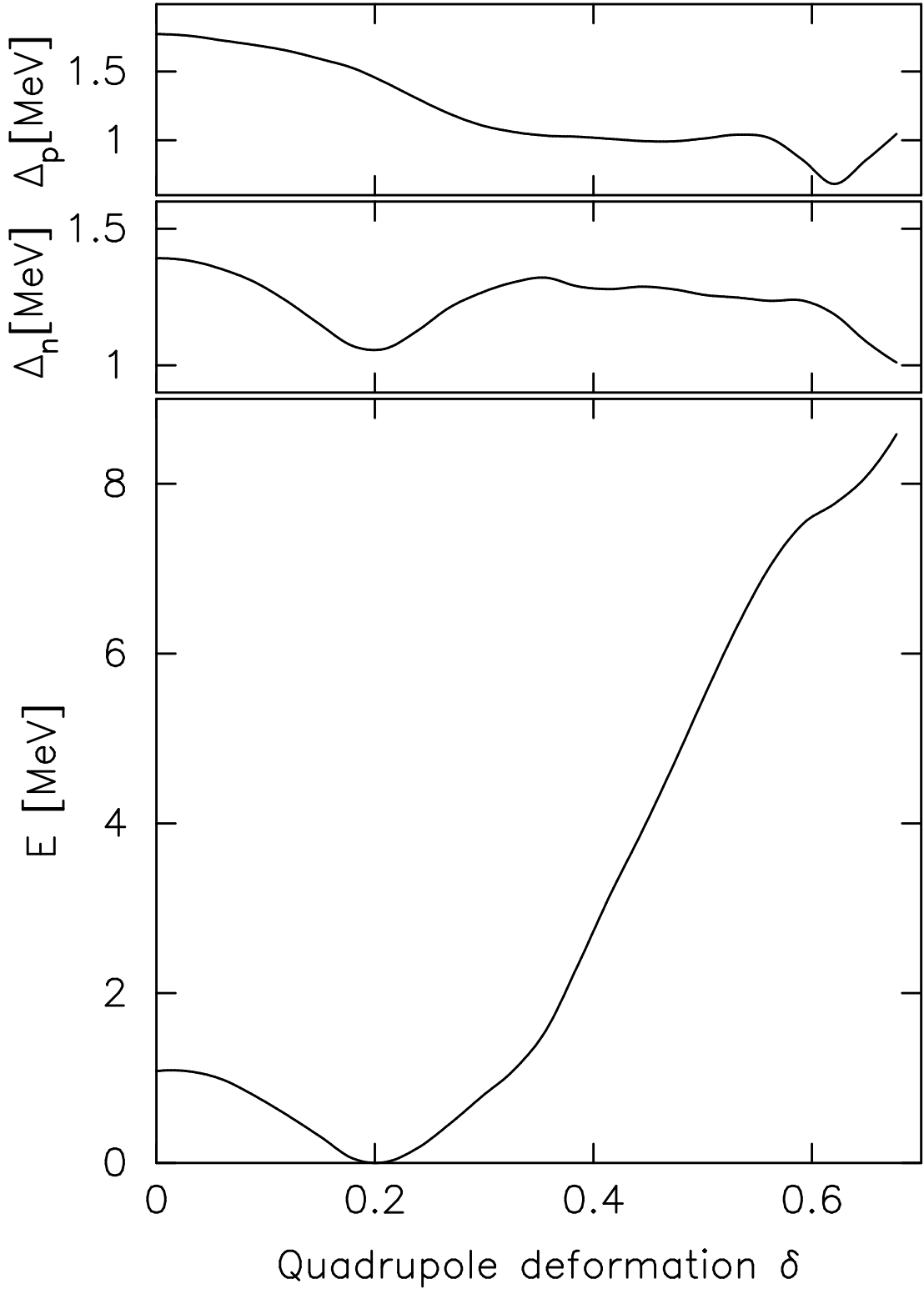
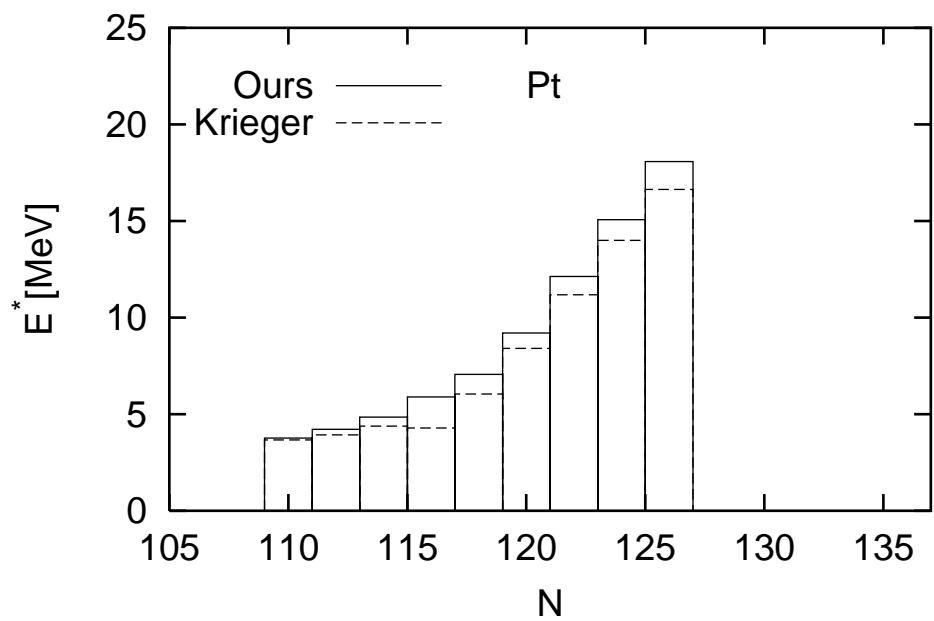
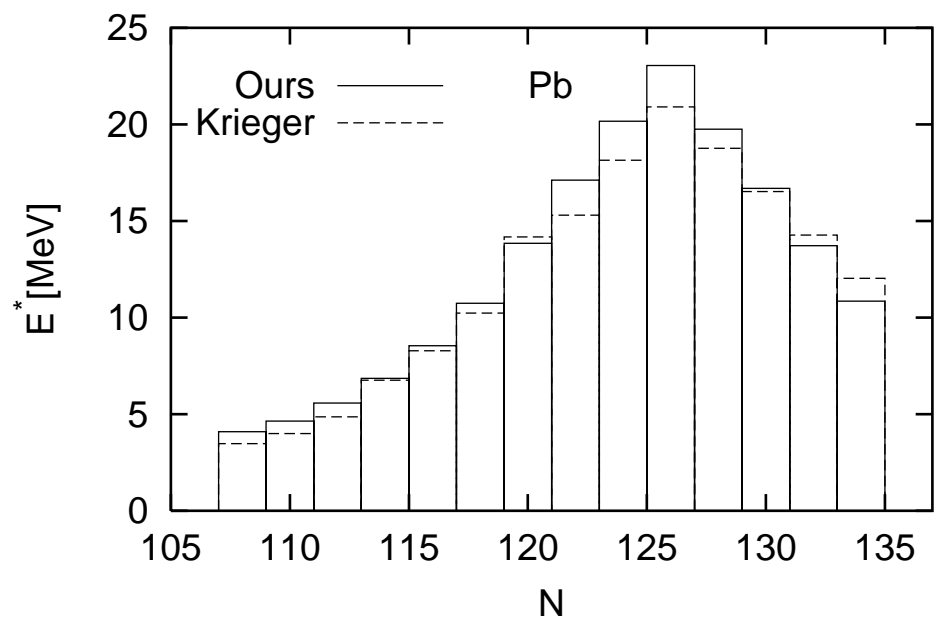
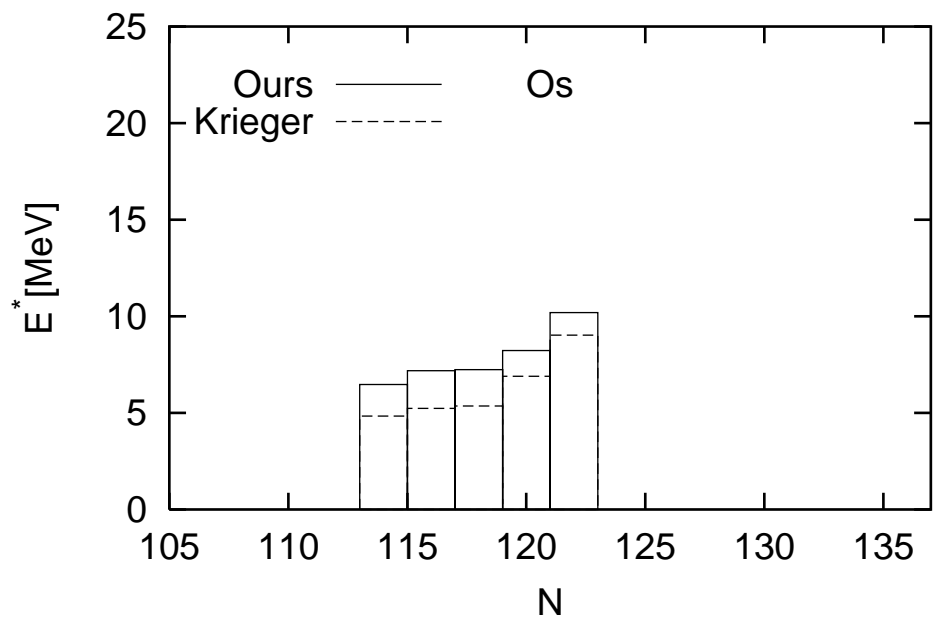
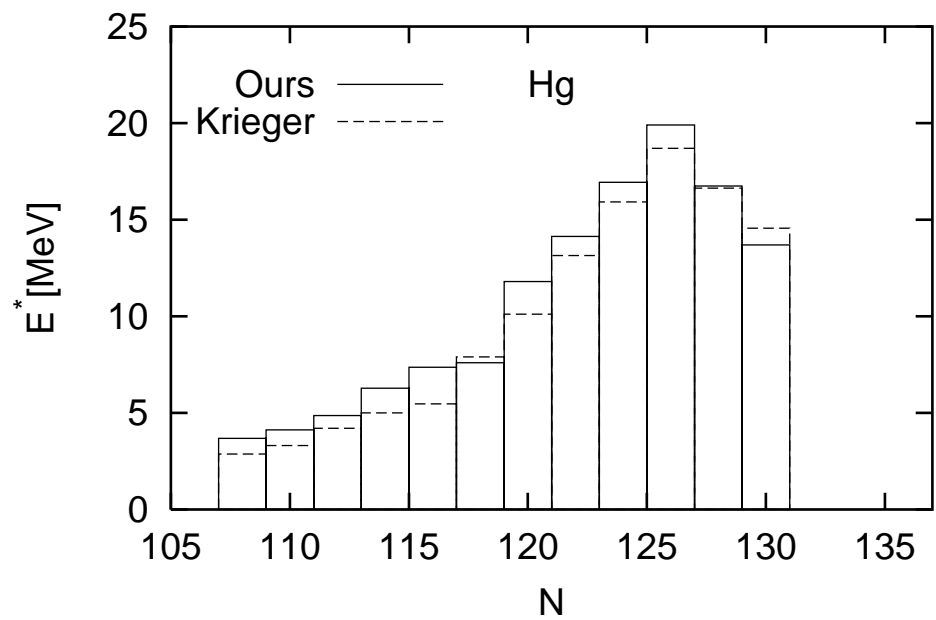
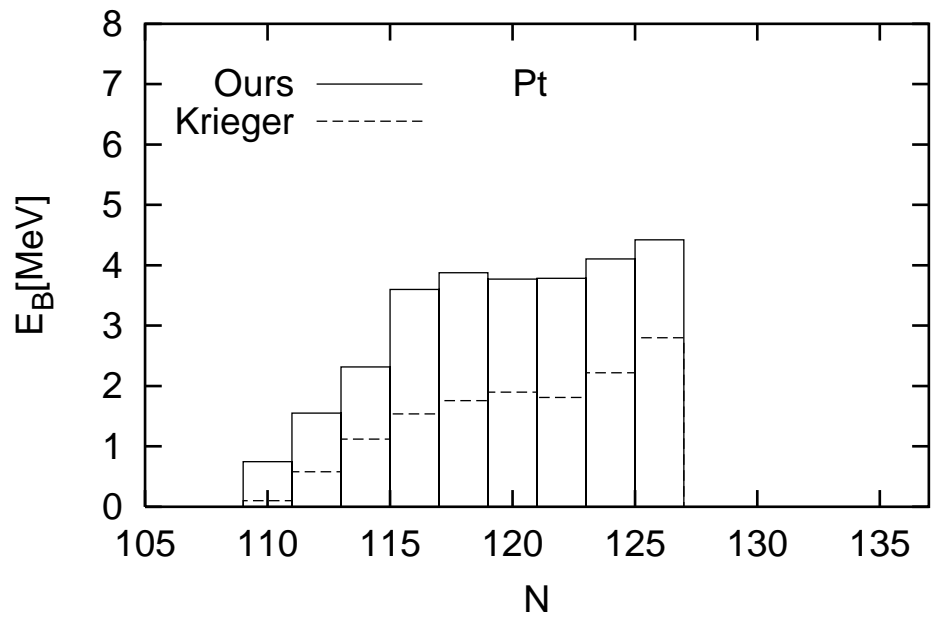
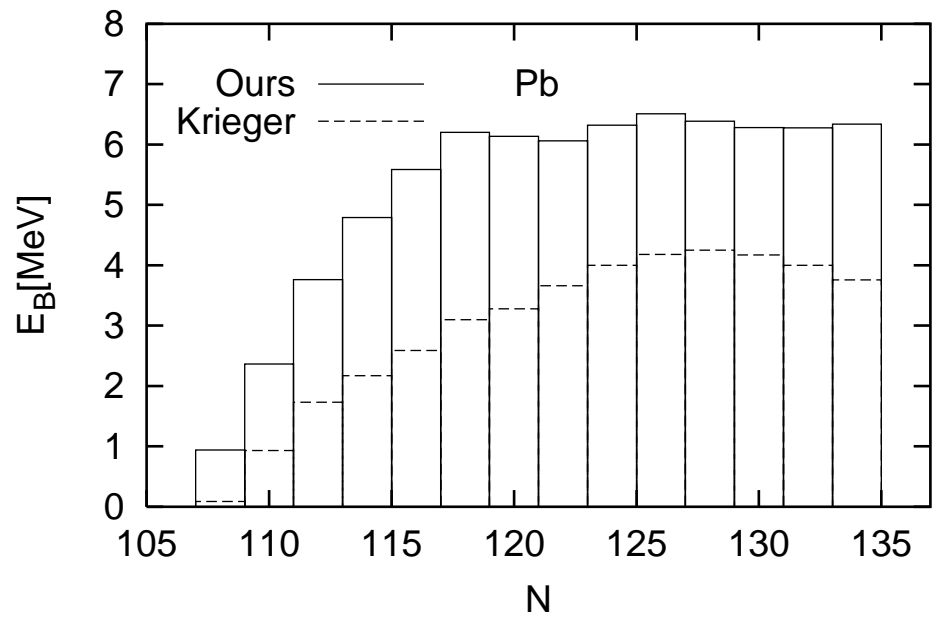
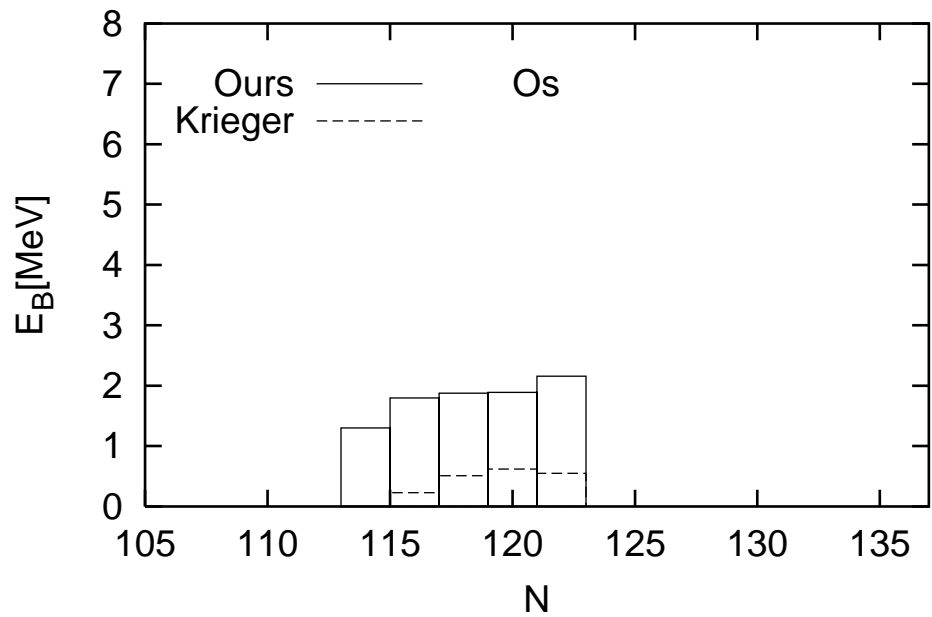
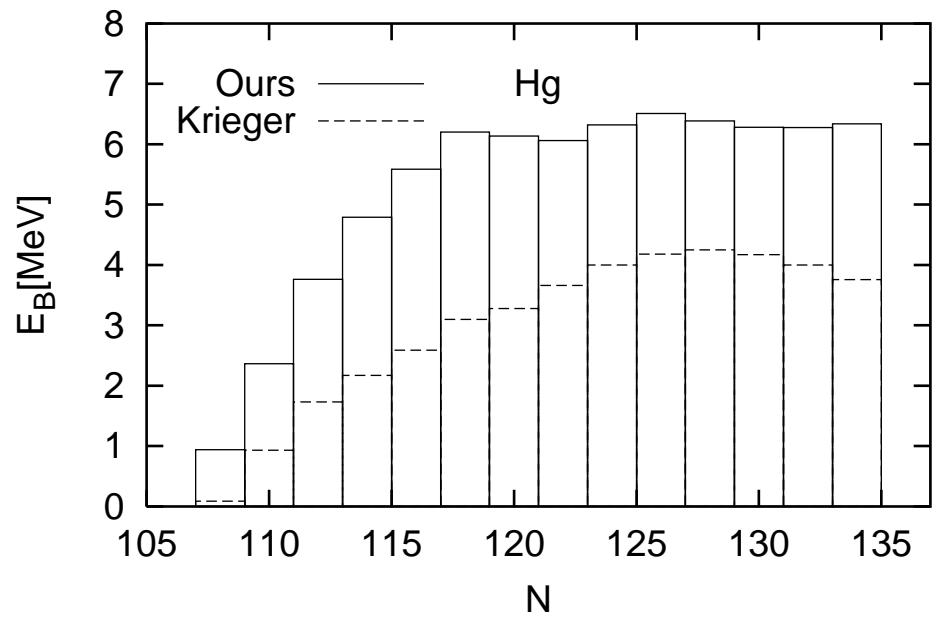
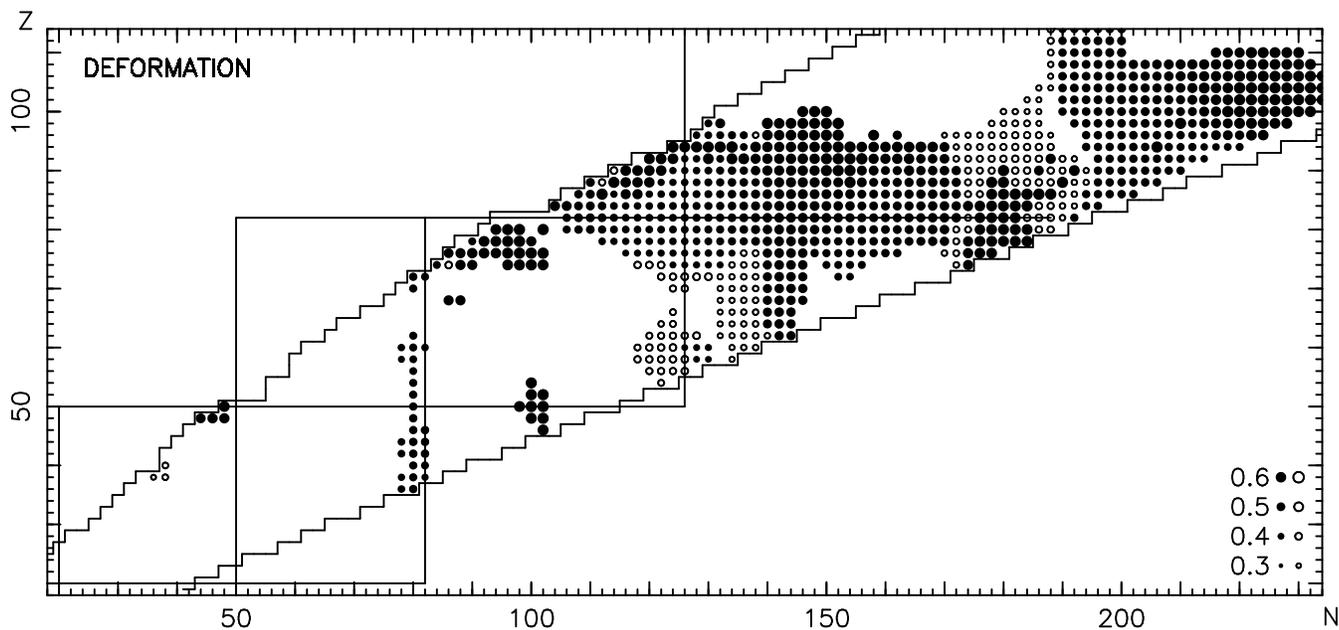
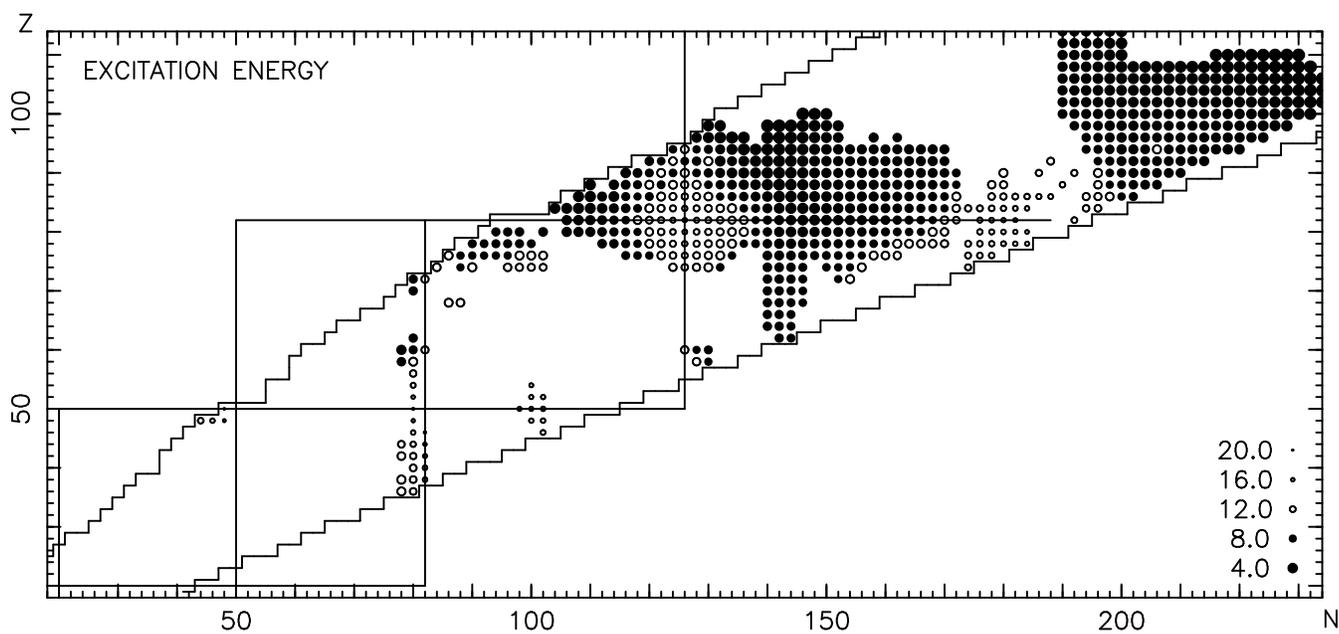
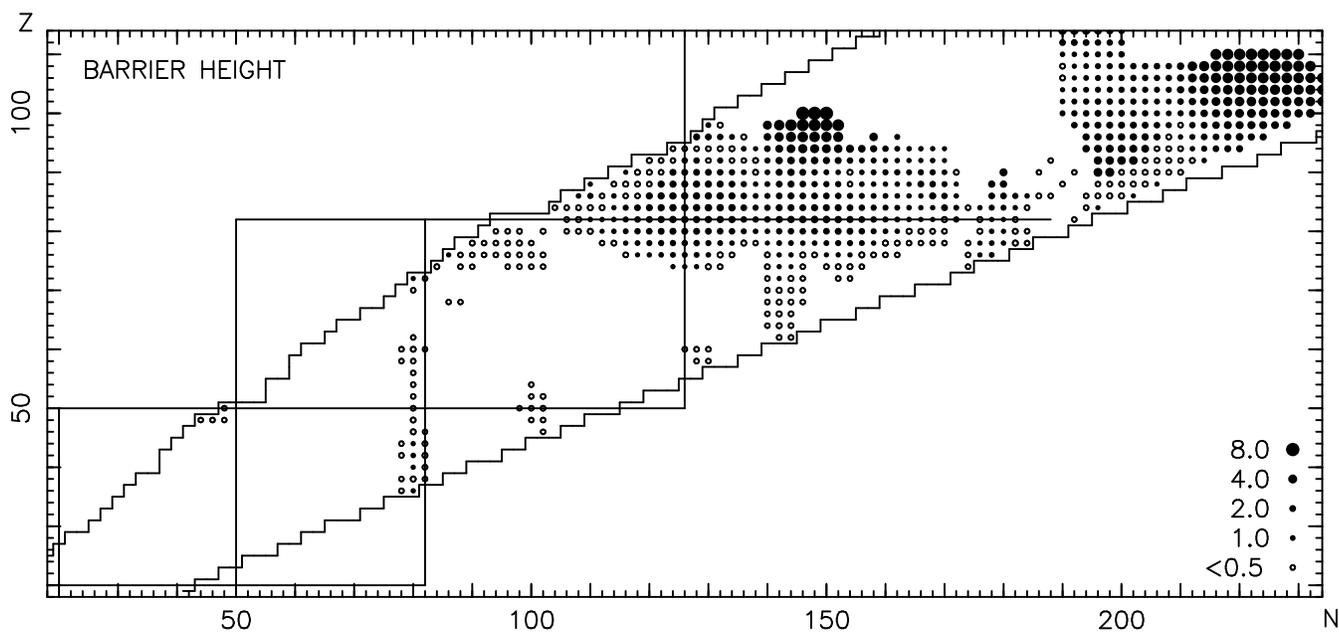
\end{document}